\documentclass[aip,apl, amsmath,amssymb, reprint]{revtex4-2}

\usepackage{graphicx}
\usepackage{dcolumn}
\usepackage{bm}
\usepackage[utf8]{inputenc}
\usepackage[T1]{fontenc}
\usepackage{mathptmx}
\usepackage{etoolbox}
\usepackage[colorlinks=true,citecolor=blue]{hyperref}
\usepackage{siunitx}

\newcommand{\edit}[1]{\textcolor{black}{#1}}

  {}{}
\begin{document}

\title{High Impedance Josephson Junction Resonators in the Transmission Line Geometry}

\author{Antti Ranni} 
\address{NanoLund and Solid State Physics, Lund University, Box 118, 22100 Lund, Sweden} 

\author{Harald Havir}
\author{Subhomoy Haldar}
\address{NanoLund and Solid State Physics, Lund University, Box 118, 22100 Lund, Sweden}

\author{Ville F. Maisi\footnote{d}}
\thanks{Authors to whom correspondence should be addressed: \hyperlink{antti.ranni@ftf.lth.se}{antti.ranni@ftf.lth.se}, \hyperlink{ville.maisi@ftf.lth.se}{ville.maisi@ftf.lth.se}}
\address{NanoLund and Solid State Physics, Lund University, Box 118, 22100 Lund, Sweden}

\date{\today}

\begin{abstract}
In this article we present an experimental study of microwave resonators made out of Josephson junctions. The junctions are embedded in a transmission line geometry so that they increase the inductance per length for the line. By comparing two devices with different input/output coupling strengths, we show that the coupling capacitors, however, add a significant amount to the total capacitance of the resonator. This makes the resonators with high coupling capacitance to act rather as lumped element resonators with inductance from the junctions and capacitance from the end sections. Based on a circuit analysis, we show that the input and output couplings of the resonator are limited to a maximum value of $\omega_r Z_0 /4 Z_r$ where $\omega_r$ is the resonance frequency and $Z_0$ and $Z_r$ are the characteristic impedances of the input/output lines and the resonator respectively.
\end{abstract}

\maketitle

High impedance resonators have obtained a significant attention in recent years. They have been used to obtain strong coherent coupling of microwave photons to semiconducting charge~\cite{stockklauser2017} and spin~\cite{landig2018, mi2018, samkharadze2018, yu2023, ungerer2023} qubits. In addition, \edit{the materials used for high impedance systems~\cite{barends2008, hutter2011, masluk2012, samkharadze2016, maleeva2018, grunhaupt2019, niepce2019, yu2021} and superinductors~\cite{bell2012, altimiras2014, peruzzo2020} are also used for realizing e.g. amplifiers~\cite{castellanos2007, macklin2015} and qubits~\cite{manucharyan2009, hazard2019, pechenezhskiy2020} thanks to their non-linearities}. The increased coupling of the \edit{high impedance resonators arises from the} added inductance \edit{giving} rise to accessing the so-called ultra-strong coupling regime \edit{for electrical dipole coupled devices}~\cite{friskKockum2019, scarlino2022}, which is predicted to give rise to new fundamental physics concepts such as electroluminesence from the ground state~\cite{cirio2016}. In this letter, we investigate high impedance ($Z_r \sim \SI{1}{k\ohm}$) resonators in a plain transmission line geometry made out of Josephson junctions. We show 
show how the \edit{capacitive} input coupling influences on the resonator properties: a modest increase of the input coupling from $\kappa_c/2\pi = 3.5$~MHz to $11$~MHz leads to a $20~\%$ reduction of the resonance frequency and the characteristic impedance. Based on these findings, we determine that - despite the ultra-strong coupling to quantum structures up to $0.6$~GHz coupling is achievable~\cite{scarlino2022} - the \edit{capacitive} input coupling is limited to $\kappa_c =  \omega_r\: Z_0/4 Z_r$ where $Z_0$ is the characteristic impedance of the input lines and $\omega_r$ the resonance frequency. For our resonators, representing typical parameter values in the recently realized experiments~\cite{masluk2012, stockklauser2017, scarlino2022}, the largest possible \edit{capacitive} coupling is limited to $\kappa_c/2\pi = 90$~MHz. This imposes a trade-off between the \edit{capacitive} input line coupling and the electric-dipole coupling to a quantum device\edit{: Using a low input coupling leads to suppressed resonator response visibility hampering the measurements via the resonator. The limitation can be, however, circumvented if the probing of the system is not made via the resonator input but for example via charge transport of the quantum system. In this case, the low input coupling can be compensated by using a larger drive signal amplitude. Note that the maximum bandwidth of the input line is however still limited by the trade-off.} The results are \edit{thus} important for designing and optimizing the high impedance resonator devices \edit{in many cases}.

\begin{figure}[t!]
    \centering
    \includegraphics[width = 0.48\textwidth]{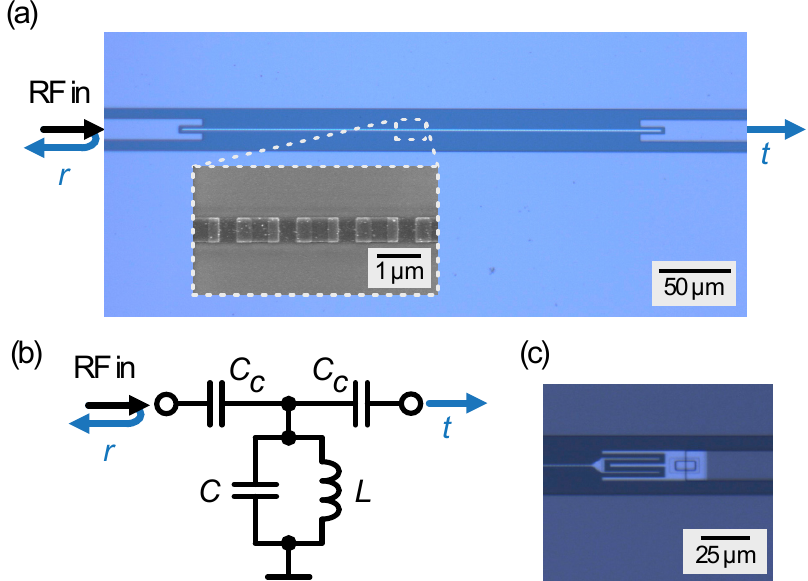}
    \caption{(a), Optical microscope image of the studied resonator with low input/output coupling strength. The thin line in the middle consists of Al-AlOx-Al Josephson junction chain forming the resonator on top of a silicon substrate with a 200 nm thick oxide layer. Thicker input and output lines to the left and right, together with the ground planes on top and bottom are made of Nb.
    An RF signal is sent in from the left port and the reflected signal with amplitude $r$ and transmission to the right port with amplitude $t$ are measured. The inset shows the Josephson junctions that form the high impedance resonator in between. (b), The electrical circuit equivalent to the resonator. The fundamental resonance mode corresponds to an LC resonator with the capacitance $C$ and inductance $L$. The capacitances $C_c$ set the input couplings. (c), Optical microscope image of the coupler geometry for the high input/output coupling resonator. \edit{Here we have used a rectangular Ti/Au contact pad with thickness 5 nm / 45 nm below the Nb and Al layers to ensure ohmic contact in the input/output lines. The larger couplers made of Al in these devices have an additional Al shadow island on top of the coupler. The island is connected by a large Josephson junction to the bottom layer. As the junction area for these junctions is $150$ times larger than for the Josephson junctions in the chain, the inductance of the junctions is vanishingly small compared to the chain inductance and the extra shadows are essentially shunted to the bottom layer.}}
    \label{fig1}
\end{figure}

Figure~\ref{fig1}~(a) shows the studied resonator. \edit{The RF input and output lines and the ground planes shown in light blue around the resonator are made of $\SI{100}{nm}$ thick Nb film sputtered in a lift-off process on a high-resistivity intrinsic silicon wafer with a $\SI{200}{nm}$ thick thermally grown silicon oxide layer. The thin Josephson junction chain between the RF lines was fabricated by a standard two-angle shadow-evaporation technique. The resist mask consisted of $\SI{1200}{nm}$ thick copolymer at bottom and $\SI{300}{nm}$ thick polymethyl methacrylate layer on top and was patterned by e-beam lithography. First, $\SI{30}{nm}$ thick Al rectangles were evaporated at an angle $\theta=\SI{20}{^{\circ}}$ to the normal of the substrate followed by an in-situ oxidation for 10 minutes in 0.36 mbar of $\mathrm{O}_2$. Then, $\SI{60}{nm}$ thick Al rectangles were evaporated at $\theta=\SI{-20}{^{\circ}}$ to connect the lower rectangles with Josephson junctions located at the overlapping areas. These Josephson junctions form a chain defining} a high impedance transmission line geometry~\cite{castellanos2007, hutter2011, masluk2012, bell2012, altimiras2014} where each Josephson junction yields an inductance $L_J = h R_T/2\pi^2 \Delta$. Here $R_T = \SI{500}{\ohm}$ is the normal state resistance of the junction, measured from similar junctions fabricated at the same processing round as the resonator, and $\Delta = \SI{200}{\micro eV}$ the superconducting gap of aluminum used as the superconductor. The junction separation of $s = \SI{1}{\micro m}$ yields the inductance per unit length of $l = L_J/s = \SI{500}{\micro H/m}$. This value is more than two orders of magnitude larger than the the permeability of vacuum $\mu_0$ which sets the typical value for non-magnetic materials. The high inductance slows down the \edit{phase velocity} of the signal~\cite{castellanos2007} such that our transmission line with the total length of $d = \SI{231}{\micro m}$ forms a $\lambda$/2-resonance mode in the $4 - 8$~GHz frequency band. \edit{For frequencies close to the resonator resonance frequency, t}his resonance mode corresponds to an equivalent LC-circuit of Fig.~\ref{fig1}~(b) with inductance $L = 2\: l\: d/\pi^2$ and capacitance $C = c\: d/2$ where $c$ is the capacitance per length~\cite{goppl2008}. 

\begin{figure}[t]
    \centering
    \includegraphics[width = 0.235\textwidth]{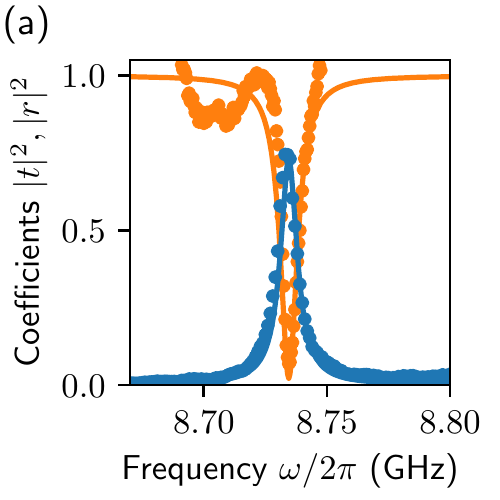}
    \includegraphics[width = 0.235\textwidth]{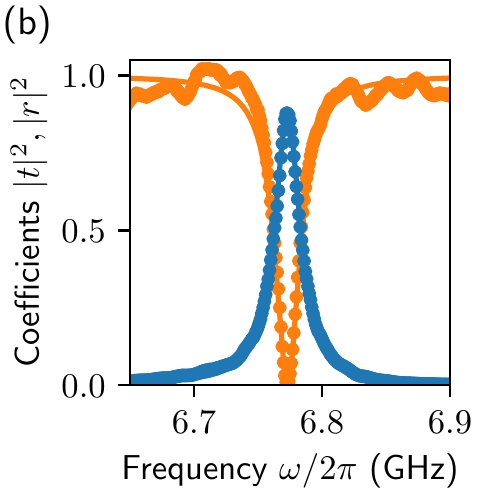}
    \caption{(a), The measured transmission $|t|^2$ (solid blue dots) and reflection $|r|^2$ (solid orange dots) coefficients as a function of the drive frequency $\omega$. The solid lines show corresponding fits to Eq.~(\ref{eq:coeffs}) with resonance frequency $\omega_r/2\pi = 8.735$~GHz, input/output coupling $\kappa_c/2\pi = 3.5$~MHz and internal losses $\kappa_i/2\pi = 1.2$~MHz. (b), The same measurements repeated for the high input/output coupling resonator. The fit parameter values are $\omega_r/2\pi = 6.773$~GHz, $\kappa_c/2\pi = 11$~MHz and $\kappa_i/2\pi = 1.3$~MHz. The measurements are done at the base temperature of 10 mK in a dilution refrigerator and the drive power applied to the left port is $-115$~dBm.}
    \label{fig2}
\end{figure}

To probe the resonator properties, we couple the resonator to standard $Z_0 = \SI{50}{\ohm}$ input and output lines with coupling capacitors $C_c$ that are identical for the input and output. Figure~\ref{fig2}~(a) presents the measured transmission $|t|^2$ and reflection $|r|^2$ coefficient as a function of the drive frequency $\omega$. A Lorentzian resonance mode at a resonance frequency of $\omega_r/2\pi = 8.735$~GHz is seen in the resonator response. The solid lines show fits to
\begin{equation}
\label{eq:coeffs}
\left\{
\begin{array}{ccc}
    |t|^2 &=& \displaystyle\frac{\kappa_c^2}{(\kappa_c+\kappa_i/2)^2 + (\omega - \omega_r)^2} \vspace{5pt}\\
    |r|^2 &=& \displaystyle\frac{(\kappa_i/2)^2 + (\omega - \omega_r)^2}{(\kappa_c+\kappa_i/2)^2 + (\omega - \omega_r)^2},
\end{array}
\right.
\end{equation}
where $\kappa_i/2\pi = 1.2$~MHz is the internal losses of the resonator and $\kappa_c/2\pi = 3.5$~MHz the input coupling~\cite{havir2023}. The input coupling and the resonator frequency are connected to the circuit elements of Fig.~\ref{fig1}~(b) as $\kappa_c = \omega_r^2\: C_c^2\: Z_0 / C_\Sigma$ and $\omega_r = 1/\sqrt{L\: C_\Sigma}$ with the total capacitance $C_\Sigma = C + 2 C_c$ of the resonator~\cite{goppl2008, havir2023}. 

In the fits of Fig.~\ref{fig2}~(a), the two input couplings $2\kappa_c$ set essentially the linewidth of the resonance and the internal losses $\kappa_i$ how high the transmission coefficient $|t|^2$ increases and reflection coefficient $|r|^2$ decreases in resonance. The resonator reaches close to the ideal values $|r| = 0$ and $|t| = 1$ of a lossless resonator, reflecting the fact that the internal losses $\kappa_i$ are significantly smaller than the couplings $\kappa_c$. \edit{Note that here the fitted value of $\kappa_i$ should not be used as precise value of the internal losses but rather as a ballpark estimate of the upper limit of the internal losses. This because the value is small compared to the total linewidth and hence is prone to large uncertainties from the background transmission and reflection calibration as well as possible spurious background transmission~\cite{rieger2023}. These uncertainties are particularly large for this case since the measured response is slightly above the $4 - 8$~GHz measurement window of our setup. Continuing now with the} Josephson junction parameter values and the \edit{other fitted values} above, we obtain $C_\Sigma = 14$~fF and $C_c = 1.4$~fF. The characteristic impedance of the resonator is then $Z_r = \sqrt{L/C_\Sigma} = \SI{1.4}{k \ohm}$. Note that here we use the characteristic impedance of the full circuit at the end points of the resonator. Hence this value describes the effective impedance value to which other circuit elements couple when connecting them to the typical maximal \edit{capacitive} coupling point at the end of the resonator with a voltage anti-node. The characteristic impedance of the Josephson junction array, $Z_\mathrm{JJ} = \sqrt{l/c} = \SI{2.4}{k\ohm}$, is close to this value but naturally doesn't depend on the coupling capacitance $C_c$ like $Z_r$ does. The capacitance per unit length of the junction array $c = 2\; C/d = 80$~pF/m matches well with similar transmission line geometries on a silicon substrate~\cite{goppl2008}. 

Next, we consider a resonator with larger input and output couplings. The Josephson junction chain of the resonator is kept the same and was fabricated in the same processing round as the low coupling device of Fig.~\ref{fig1}~(a). The input and output couplings are made larger by using an inter-digitized finger geometry shown in Fig.~\ref{fig1}~(c). The resulting response is presented in Fig.~\ref{fig2}~(b) together with the fits yielding $\omega_r/2\pi = 6.773$~GHz, $\kappa_c/2\pi = 11$~MHz and $\kappa_i/2\pi = 1.3$~MHz. \edit{Again, the small value of $\kappa_i$ should be taken as an indicative one for the upper limit of the internal losses.} With the same inductance $L$ from the Josephson junction chain, the corresponding circuit parameter values are thus $C_\Sigma = 23$~fF and $C_c = 4.1$~fF resulting in $Z_r = \sqrt{L/C_\Sigma} = \SI{1.0}{k \ohm}$. 

When comparing the two resonator responses, we see that despite the input/output coupling is increased merely to $\kappa_c/2\pi = 11$~MHz, the resonance frequency drops by $2$~GHz. This shift arises as the coupling capacitances $2C_c$ contribute to nearly half of the total capacitance $C_\Sigma$ for the high coupling resonator. Correspondingly, the characteristic impedance $Z_r$ drops by the equal amount of $20 \%$ due to the significant added capacitance. On the other hand, the nearly unaltered internal losses $\kappa_i$ and the Josephson junction chain capacitance of $C = 11$~fF for the low coupling resonator versus $C = 15$~fF for the high coupling resonator yield good consistency checks for our model. The extra capacitance of $4$~fF in the high coupler resonator is likely due to the increased self capacitance and extra capacitance to ground from the inter-digitized couplers. Indeed, the extra length of $\Delta l = \SI{40}{\micro m}$ from the couplers corresponds to a capacitance of $c\; \Delta l = 3$~fF.

\begin{figure}[t]
    \centering
    \includegraphics[width = 0.48\textwidth]{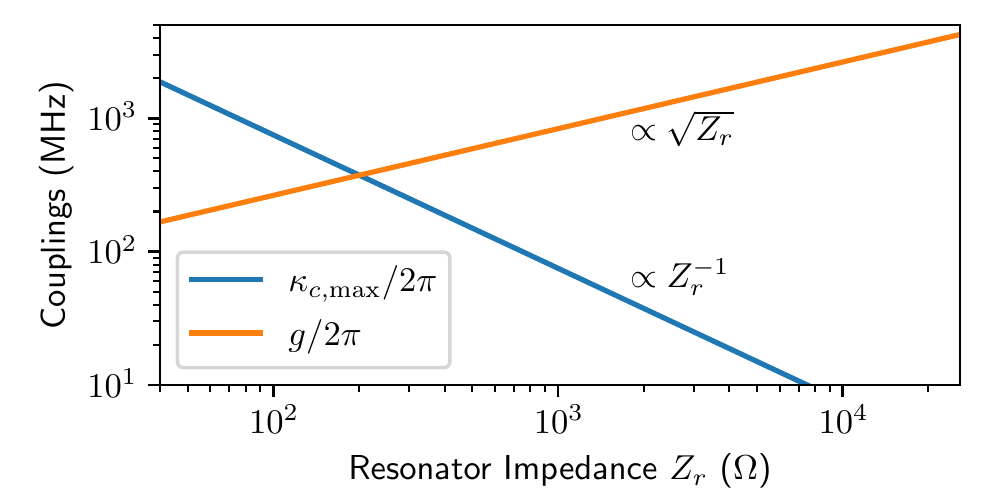}
    \caption{\edit{The largest input coupling $\kappa_{c,\mathrm{max}}$ of Eq.~(\ref{eq:kappacmax}) and dipole coupling $g$ for a double quantum dot charge qubit at sweet spot with zero detuning~\cite{childress2004}. The parameter values used for the plots are $Z_0 = \SI{50}{\Omega}$, $\omega_r/2\pi = 6$~GHz, and the charge qubit is taken with maximal capacitive coupling with the lever arm $\nu = 1$.}}
    \label{fig3}
\end{figure}

The above comparison of the two resonators yields two important points for high impedance resonator designing: 1) Even with the rather small input couplings, one needs to pay extra attention to the added total capacitance that may otherwise shift the resonance frequency uncontrollably far away from the desired value, as well as lower the resonator impedance $Z_r$. This is in stark contrast to the low impedance $\SI{50}{\ohm}$ resonators where the resonance frequency shift is comparable to the coupling $\kappa_c$, see e.g. Ref.~\citealp{goppl2008}. 2) The significant contribution of the coupling capacitances to the total capacitance limits the highest attainable input/output coupling value $\kappa_{c,\mathrm{max}}$. We obtain this largest coupling by taking the coupling capacitance to dominate the total capacitance, i.e. $C_c \gg C$, which yields $C_\Sigma = 2C_c$ and the highest coupling as
\begin{equation}
\label{eq:kappacmax}
    \kappa_{c,\mathrm{max}} = \frac{\omega_r^2\: C_c^2\: Z_0 } {C_\Sigma} = \frac{\omega_r^2\: C_\Sigma\: Z_0 }{4} = \frac{Z_0 }{4 Z_r} \omega_r.
\end{equation}
In the last equality, we used the relation $\omega_r Z_r = 1/C_\Sigma$. For our resonators, exemplary for typical ones used in the experiments, we have $\kappa_{c,\mathrm{max}}/2\pi = 90$~MHz with $Z_r = \SI{1}{k\ohm}$ and $\omega_r/2\pi = 7$~GHz. \edit{By using a single port resonator, the lowest total capacitance would be $C_\Sigma = C_c$ instead, and the highest possible coupling increases to $\kappa_{c,\mathrm{max}} = \omega_r\: Z_0/Z_r$.} Interestingly, these kind of resonators have been shown to yield coherent coupling in the strong and ultra-strong coupling regime for interactions between the microwave photons \edit{- or more precisely plasmons~\cite{mooij1985} -}  in the resonator and a charge degree of freedom in a semiconducting quantum dots~\cite{scarlino2022}. In these experiments, the dipole coupling value reached up to $g/2\pi = 600$~MHz 
\edit{with $Z_r = \SI{4}{k \Omega}$ and $\omega_r/2\pi = \SI{5.6}{GHz}$. The largest input coupling in this device is limited then to $\kappa_{c,\mathrm{max}} = \SI{70}{MHz}$, an order of magnitude lower than $g/2\pi$.}
\edit{Figure \ref{fig3} summarizes the trade-off between $\kappa_{c,\mathrm{max}}$ and the dipole coupling $g = \omega_r \nu \sqrt{Z_r/2 R_Q}$ of a semiconductor charge qubit~\cite{childress2004} serving as an exemplary system to which the resonators couple to. Here $\nu$ is the capacitive coupling constant between the qubit and the resonator and $R_Q = h/e^2$ the resistance quantum. We see that for $Z_r > \SI{1}{k\Omega}$, the input coupling will be significantly smaller than than $g$, even for the more modest charge qubit couplings of the order of $\nu \sim 0.1$, typical in many experiments~\cite{frey2012, scarlino2022, mi2017, khan2021, haldar2023}.}

To recover the low impedance result connecting the frequency shift to the input coupling directly, the same characteristic impedance could be used for both the resonator and the input and output lines with $Z_0 = Z_r$. Forming the input and output lines with the Josephson junction arrays approach would be possible for our resonators without any further processing steps. However, an additional consideration would be needed for connecting the lines further to the usually used standard $\SI{50}{\ohm}$ cabling. A "tapering" of the characteristic impedance by increasing the junction separation $s$ gradually along the line could be used. The high kinetic inductance approaches~\cite{barends2008, samkharadze2016, maleeva2018, niepce2019} would offer also a viable option here as the characteristic impedance could be changed by widening the input and output lines gradually with the tapering. \edit{Using an inductive coupling scheme instead of the capacitive one may also provide a way to obtain couplings larger than the capacitive limit studied here.}

In conclusion, we have measured the transmission and reflection coefficient response of high impedance transmission line resonators made out of Josephson junctions. By varying the capacitive input coupling, we showed that the capacitance added by the couplers has a much stronger effect on the high impedance resonators in comparison to the low impedance ones. The capacitance both reduces the resonance frequency significantly as well as reduces the impedance level. We also showed that the maximum input/output coupling is reduced and limited to $\kappa_{c,\mathrm{max}} = \omega_r Z_0 /4 Z_r$, despite the dipole coupling is increased to an order of magnitude larger values for the typical resonator impedances values of the order of $Z_r \sim \SI{1}{k \ohm}$ used in these experiments.

We thank A. Baumgartner, A. Pally, P. Scarlino, C. Schönenberger, C. Thelander and J. Ungerer for fruitful discussions, and Swedish Research Council (Dnr 2019-04111), the Foundational Questions Institute, a donor advised fund of Silicon Valley Community Foundation (grant number FQXi-IAF19-07) and NanoLund for financial support.

%

\end{document}